# Analytical Study of KOH Wet Etch Surface Passivation for III-Nitride Micropillars

Matthew Seitz,[1,5*] Jacob Boisvere,[1] Bryan Melanson,[1] John Wyatt Morrell,[2,3] Nithil Harris Manimaran,[1] Ke Xu,[1,2,3] Jing Zhang[4,5*]

[1]Microsystems Engineering, Rochester Institute of Technology, Rochester, New York, 14623, USA
[2]School of Chemistry and Material Science, Rochester Institute of Technology, Rochester, New York, 14623, USA
[3]School of Physics and Astronomy, Rochester Institute of Technology, Rochester, New York, 14623, USA
[4]Department of Electrical and Microelectronic Engineering, Rochester Institute of Technology, Rochester, New York 14623, USA
[5]Lead Contacts
*Correspondence: ms3897@rit.edu, jzeme@rit.edu

**SUMMARY**
III-Nitride micropillar structures show great promise for applications in micro light-emitting diodes and vertical power transistors due to their excellent scalability and outstanding electrical properties. Typically, III-Nitride micropillars are fabricated through a top-down approach using reactive ion etch which leads to roughened, non-vertical sidewalls that results in significant performance degradation. Thus, it is essential to remove this plasma etch induced surface damage. Here, we show that potassium hydroxide (KOH) acts as a crystallographic etchant for III-Nitride micropillars, preferentially exposing the vertical $<1\bar{1}00>$ m-plane, and effectively removing dry etch damage and reducing the structure diameter at up to 36.6 nm/min. Both KOH solution temperature and concentration have a dramatic effect on this wet etch progression. We found that a solution of 20% AZ400K (2% KOH) at 90°C is effective at producing smooth, highly vertical sidewalls with RMS surface roughness as low as 2.59 nm, ideal for high-performance electronic and optoelectronic devices.

III-Nitride, microstructures, surface passivation, plasma etch, wet etch.

## INTRODUCTION

The search for ever smaller electronic and optoelectronic devices has led to numerous innovations. Large planar devices have become micropillars and nanowires,[1] allowing aggressive scaling for enhanced functionalities. This paves the way for higher density integrated transistors for more powerful computing,[2] and brighter micro light-emitting diodes (μLEDs) for higher resolution displays for augmented reality (AR) and virtual reality (VR),[3,4] as well as visible light communication.[5] III-Nitrides including gallium nitride (GaN) offers additional innovations in power electronics and optoelectronics and is a significant research interest due to its wide, direct bandgap, as well as its high electron mobility, and excellent thermal properties.[6,7]

III-Nitride micropillar or nanowire based electronic and optoelectronic devices can be fabricated through either bottom-up or top-down strategies. In a bottom-up approach, device structures are grown on a substrate through molecular beam epitaxy (MBE)[8] or metalorganic chemical vapor deposition (MOCVD).[9] These growth approaches can result in highly dense forests of individual micopillars or nanowires; however, this unconstrained growth can lead to non-uniformity and coalescence of adjacent pillars or wires.[10] Through the use of a patterned mask on the growth surface, micropillar or nanowire nucleation sites can be intentionally defined, forming ordered arrays of micropillars or nanowires rather than disordered forests.[11] However, non-uniformities between nanowires still limit the effectiveness of these bottom-up approaches.

Epitaxial growth can instead be performed at the wafer scale, enabling a top-down fabrication strategy. In this case, similar MBE or MOCVD epitaxial growth produces layered thin films that can be fabricated into a range of devices and geometries through photolithographic patterning and etching.[12] The use of well-established lithographic processing also offers an extremely high degree of uniformity and control over the placement, size, and geometry of the resulting structures compared to the probabilistic growth of bottom-up fabrication. Etching is typically accomplished using reactive ion etching (RIE) with inductively coupled plasma (ICP) using a $Cl_2$/Ar gas mixture.[12] However, this RIE process also causes crystalline damage in the form of dangling bonds and surface roughness, which form current leakage paths which can severely limit device performance as well as cause non-radiative recombination in optoelectronic devices.[13,14]

Several fabrication challenges remain unsolved in this top-down fabrication methodology. GaN is a highly chemically inert material system and the effectiveness of wet etching processes are very limited.[7] Due to this, virtually all top-down GaN fabrication involves one or more dry etch process steps. Dry etching is effective at removing material to form device structures and is a highly mature approach. However, these processes also cause damage to the newly exposed surfaces and typically lead to the formation of non-vertical sidewalls.[15,16] These defects present challenges for both device fabrication and performance. Slanted sidewalls can complicate subsequent deposition steps, and surface damage contributes to non-radiative recombination which significantly reduces the optical output and degrades device efficiency. To optimize device performance, the dry etch damaged III-Nitride micropillar sidewall surface must be fully removed. Therefore, here, we propose an analytical study on the use of KOH-based wet etch that can effectively passivate vertical III-N nanowires and micropillars. This expands upon our previous work[17] which only qualitatively considered etchant temperature. In this work, we demonstrate a KOH-based wet etch that can effectively passivate vertical III-N nanowires and micropillars and investigate the effects of etchant concentration, temperature, and the presence of an insoluble metal mask. The effects of the different etch conditions are analyzed comprehensively, and III-Nitride micropillar sidewall surface roughness was measured and compared from our study.

## RESULTS AND DISCUSSION
### Background and theory

Several approaches have been investigated to remove or repair dry etch damage and improve overall device performance and efficiency for III-Nitride nano- and micro-structures. Passivation through atomic layer deposition (ALD) of dielectrics or treatment with sulfide solutions have been shown to minimize surface recombination, improving device performance and efficiency.[18,19] However, this requires additional time consuming and expensive processing, and adds additional material to the device surface that can affect the optical properties of the device. Additionally, achieving the necessary fully conformal passivation coating presents further challenges. As an alternative, dilute KOH solutions have also been shown to improve dry etched GaN surfaces for edge-emitting lasers.[15] Through this KOH method, damaged material at the surface is etched away, removing the current leakage paths and non-radiative recombination centers, thus improving overall device performance for devices such as μLEDs.[17] However, no comprehensive study has been reported yet for vertical III-Nitride micropillars or nanowires. Thus, in this work, AZ400k, a common commercially available photoresist developer is used as the KOH source. This developer contains 2 wt% KOH in solution and is further diluted to 20%, 40%, or 60% AZ400k with deionized water, resulting in an etchant solution with 0.4%, 0.8%, or 1.2% KOH.

This dilute aqueous KOH solution etches GaN-based structures through a two-step process. KOH first acts as a catalyst for the oxidation of GaN by hydroxide ions (OH$^-$), forming gallium oxide and ammonia. Subsequently, the KOH acts as a solvent, dissolving and removing the newly-formed gallium oxide. This etch progresses quickly for N-polar GaN crystals, but significantly slower for Ga-polar crystals due to repulsion between the OH$^-$ ions and the GaN internal polarization field.[7,20] Due to these interactions, KOH acts as a crystallographic etchant for GaN, preferentially exposing polar planes such as the $<1\bar{1}00>$ plane while leaving the Ga-polar <0001> plane untouched. This makes KOH an extremely effective means to form vertical structures with highly smoothed sidewalls, as well as to reduce the diameter of etched features without affecting the height of the structure.[17] In this work, we examine the effects of KOH concentration, solution temperature, and the presence of a metal mask on the progression of this wet etch. The presence of a metal mask mimics a self-aligned micropillar fabrication process where deposition of electrical contacts after dry etching becomes impractical due to the small size of the devices.[21]

### Experimental

Two sets of micropillars were prepared from an AlGaN/GaN wafer composed of a 0.46 μm layer of Al$_{0.19}$Ga$_{0.81}$N atop a 4.7 μm layer of GaN. As shown in Figure 1, fabrication of the first set of micropillars begins by first coating the sample with 500 nm SiO$_2$ via plasma enhanced chemical vapor deposition (PECVD) of tetraethyl orthosilicate (TEOS). Samples were coated with lift-off resist and positive photoresist, then patterned via direct-write lithography. A 150 nm layer of nickel (Ni) was thermally evaporated, deposited and lifted-off. This Ni was first used to mask a fluorine-based plasma etch of the SiO$_2$, revealing the underlying AlGaN surface. The remaining Ni/SiO$_2$ mask was then used to mask a Cl$_2$/Ar etch to form the micropillars themselves. Immersion in HF-containing buffered oxide etch solution dissolved the SiO$_2$, effectively removing the Ni/SiO$_2$ mask. This culminated in the formation of a sparse array of 2 μm tall, 2 μm diameter micropillars with flared bases and sidewalls roughened by the dry etching process.

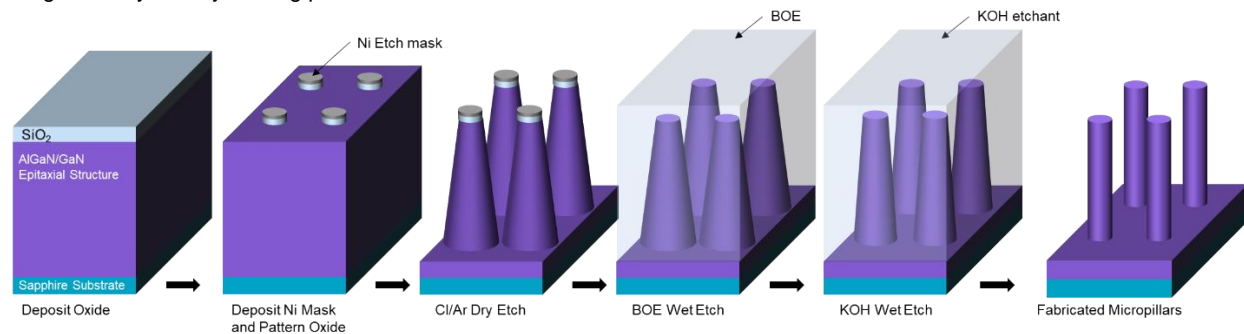

**Figure 1. Fabrication process flow for unmasked micropillars**
Schematic of the micropillar fabrication process consisting of SiO2 deposition, deposition and lift off of a Ni etch mask and its use during dry etching of first the SiO2, then the AlGaN/GaN epitaxial structure. Following dry etching, the remaining SiO2 and metal masks are removed by immersion in BOE and the micropillars undergo a novel KOH-based wet etch to form smooth, vertical sidewalls.

The second set of micropillars followed a similar fabrication process but omitted the SiO$_2$ interlayer. As shown in Figure 2, after coating the sample with lift-off resist then positive photoresist and identical patterning via direct-write lithography, the same 150 nm layer of Ni was thermally evaporated and deposited directly on the AlGaN surface, and was lifted-off. This Ni was used as an etch mask for an identical Cl$_2$/Ar etch, also resulting in similar sparse arrays of flared-base micropillars, with a Ni mask still covering the top surface.

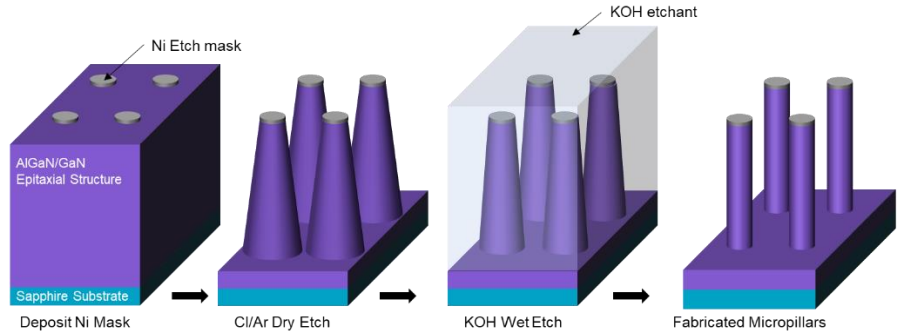

**Figure 2. Fabrication process flow for Ni masked micropillars**
Schematic of the micropillar fabrication process consisting of deposition and lift-off of a Ni etch mask, dry etching of the AlGaN/GaN epitaxial structure, and final wet etching using a novel KOH-based wet etch.

After collecting baseline images of the micropillars through scanning electron microscopy (SEM) before any wet etching, pairs of samples consisting of both masked and unmasked AlGaN/GaN micropillars received a series of wet etches under a range of conditions. Each etching step was performed in 20-minute increments, allowing for observation of the progression of the wet etch over time. After each wet etch cycle, the samples were imaged via SEM. All samples received three rounds of wet etching, for a total of 60 minutes etch time.

Wet etching was performed using MicroChemicals AZ400K, a KOH-containing photoresist developer at a range of concentrations and solution temperatures. AZ400K was diluted with deionized water to concentrations of 20%, 40%, and 60% to serve as etch solution. These solutions were then heated to 70 °C, 80 °C, and 90 °C under constant stirring to perform the etch. Pairs of samples, one with a Ni mask and one without, were placed in dipper baskets and etched together. After each 20-minute etch cycle, the pair of samples were removed, rinsed in deionized water, blown dry, and imaged via SEM. Following the completion of wet etching, the micropillar sidewall surface roughness was measured via atomic force microscopy (AFM).

### Results

One of the challenges of studying crystallographic etching processes is the need to carefully and precisely align features with the preferentially exposed crystal planes. For these experiments, we selected circular micropillars to avoid the need for this precise alignment and instead rely on the inherent symmetry of the micropillars to allow the structures to effectively self-align crystallographically, as shown in Figure 3. This enables us to analyze and observe the progression and effectiveness of the wet etching process while avoiding the possible unintended introduction of additional sidewall roughness due to crystallographic misalignment.

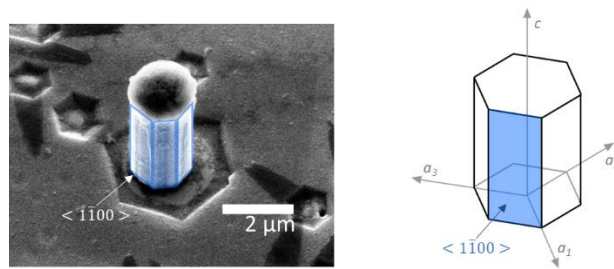

**Figure 3. Micropillar following wet etching showing smooth crystal planes.**
SEM image of a Ni-capped micropillar after undergoing wet etching in KOH-based solution and a schematic diagram of the crystal planes exposed during the wet etch.

After dry etching in $Cl_2$/Ar, the micropillars were formed with slanted, non-vertical sidewalls that were scored with small vertical grooves, indicative of dry etch damage. This damage is steadily removed during the wet etching process but progresses differently for masked and unmasked structures. For structures without a Ni capping layer, the top-most AlGaN layer etches more quickly than the lower GaN layer. As the wet etch progresses, the diameter of the AlGaN layer steadily reduces, exposing more and more of the c-plane GaN surface at the AlGaN-GaN interface. At this newly exposed GaN surface, the KOH etch begins etching downward, revealing the vertical $< 1\bar{1}00 >$ plane. However, as this downward etch occurs, the upper AlGaN layer also continues to etch radially inward and further reduce in diameter, exposing additional GaN at the interface. This newly exposed GaN also begins to etch downward, resulting in the formation of numerous steps and terraces down the side of the micropillar, as shown in Figures 4(a-d).

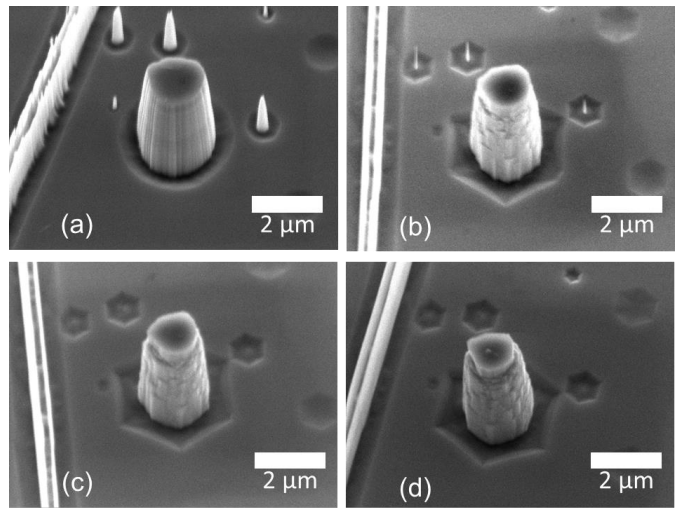

**Figure 4. Wet etch progression for micropillars without a Ni mask**
(a) prior to wet etch, after a (b) 20 minute, (c) 40 minute, and (d) 60 minute wet etch

Structures that are capped with a layer of Ni show a very different etch progression. The inclusion of a completely insoluble mask atop the structure dramatically reduces the radially-inward etch rate of the top AlGaN layer. This slower etch rate enables the formation of smooth, vertical sidewalls without the terraces in the lower GaN layer that are seen in samples without the Ni mask. Since additional GaN is exposed at a much slower rate at the AlGaN-GaN interface, the wet etch can expose the same $< 1\bar{1}00 >$ plane along the full height of the pillar, producing a uniform surface, shown in Figure 5(a-d).

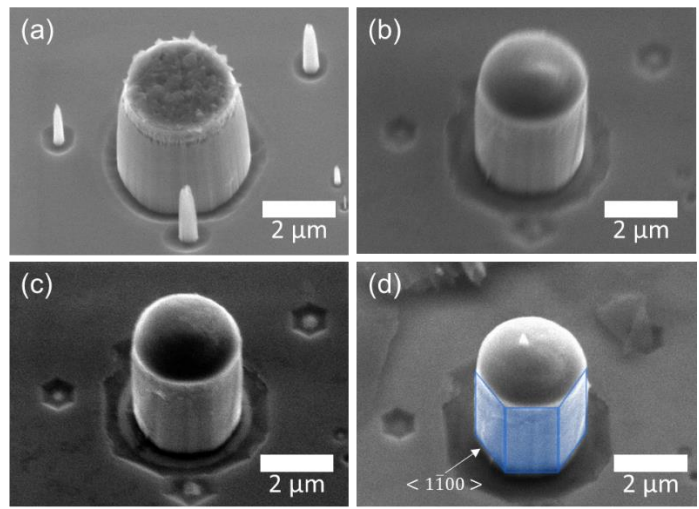

**Figure 5. Wet etch progression for micropillars with a Ni mask**
(a) prior to wet etch, after a (b) 20 minute, (c) 40 minute, and (d) 60 minute wet etch

In all the cases, we observe a reduction in III-Nitride micropillar diameter as the wet etch progresses. This etch rate is highest at the beginning of the etch as the outer surface of the pillar, damaged by the initial dry etch, is quickly removed. The undamaged AlGaN/GaN etches more slowly, leading to an overall reduction in etch rate with time. As shown in Figure 6(a), we observe an etch rate of 32.3 nm/min for the sample with a Ni etch mask during the first 20 minute etch cycle in an 80 °C solution of 40% AZ400K. This etch rate falls to only 1.9 nm/min during the third and final 20 minute etch cycle. These etch rates are slightly slower than the etch rates observed for the sample without a Ni mask. When exposed to identical etch conditions, we observe an etch rate of 36.6 nm/min for the unmasked sample during the first 20 minute etch cycle. This etch rate falls to 5.2 nm/min during the final 20 minute etch cycle. Both increased solution temperature and increased AZ400K concentration leads to a more aggressive, faster etch. When exposed to an 80 °C etch solution of 60% AZ400K, we observe an etch rate of 24.7 nm/min for an unmasked sample. When an identical sample is exposed to the same etch solution at 70 °C, this etch rate falls to just 3.4 nm/min, as shown in Figure 6(b). Solution concentration has a similar, but less dramatic effect on observed etch rate. As shown in Figure 6(c), during the first etch cycle, identical, Ni-capped samples show an etch rate of 32.1 nm/min in a 40% AZ400K solution at 90 °C compared to an etch rate of 29.3 nm/min in a 20% AZ400K solution. The difference in etch rate becomes larger with time. During the final 20 minute etch cycle, we observe an etch rate of 2.1 nm/min using a 40% solution, but only observe an etch rate of 0.7 nm/min using a 20% solution. These results suggest that the wet etch rate of crystalline III-Nitrides, undamaged by prior dry etching, is much more sensitive to etch conditions. Conversely, III-Nitrides which had been damaged by a prior dry etch show a more consistent wet etch rate in KOH solution across a range of etch conditions.

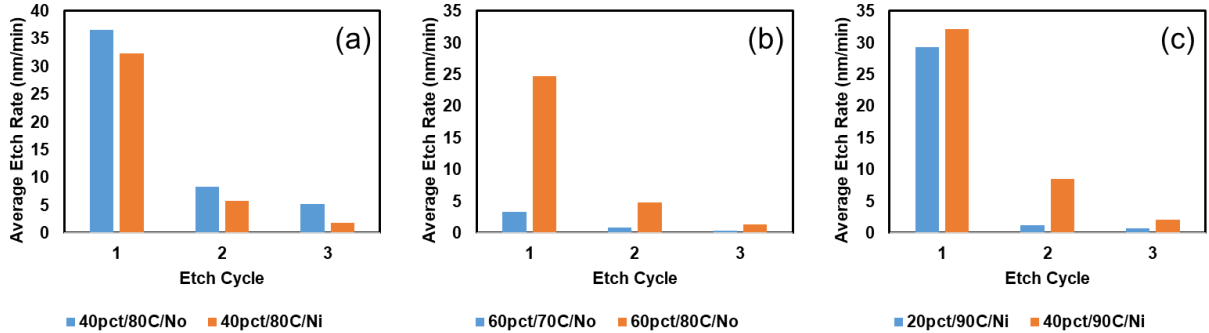

**Figure 6. Etch rate comparisons for micropillars**
(a) with and without a Ni mask, (b) using different etch solution temperatures, (c) using different etch solution concentrations.

Based on the SEM images collected following etching under different conditions, the etching of Ni-capped micropillars in a solution of 20% AZ400K heated to 90 °C yielded the smoothest sidewalls. Post-etch SEM images show that the presence of a Ni etch mask prevents the formation of terraces on structure sidewalls and that lower etchant concentrations were sufficient to produce the desired crystallographic etch. Additional AFM measurements were taken using micropillars which were mechanically removed from their substrate before and after wet etching under these optimized conditions, shown in Figure 7. These measurements indicate that RMS surface roughness reduces from 33.09 nm prior to wet etching to 2.59 nm after 60 minutes wet etching in KOH solution. This significantly reduced RMS is comparable to state-of-the-art epitaxially grown GaN thin film with typical RMS of 2.5 nm,[22] clear quantitative evidence of the removal of dangling bonds and surface damage, improving device performance and reducing current leakage.

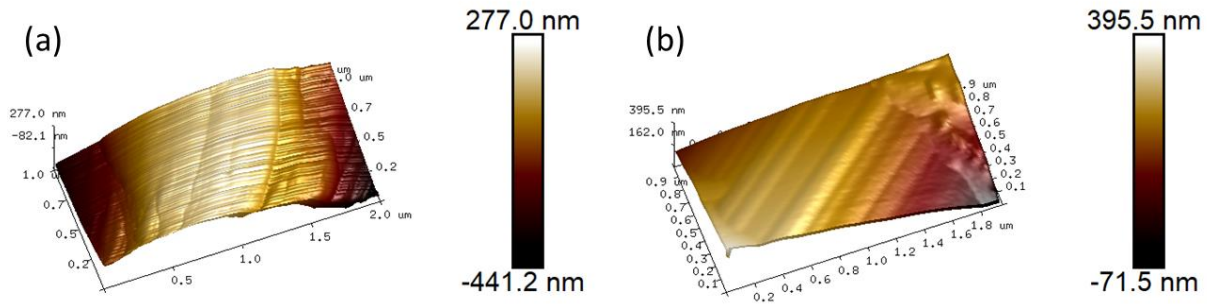

**Figure 7. AFM measurement results**
(a) before wet etching in KOH solution, (b) after 60 minutes wet etching.

## Conclusions

In summary, we have explored the effects of KOH solution concentration and temperature as well as the effects of a metal hard mask on the wet etching of vertical III-Nitride micropillar structures. Solution temperature plays a major role in the observed etch rate, increasing solution temperature from 70 °C to 80 °C causes the observed etch rate to increase from 3.4 nm/min to 24.7 nm/min. Changes in solution concentration have a more minor effect on the initial etch rate, increasing from 20% AZ400K to 40% led to an etch rate increase of only 2.8 nm/min, an increase of approximately 10%. However, as the etch progresses and material damaged by prior dry etching is removed, the etch rate of the underlying undamaged GaN is more sensitive to solution concentration. During the final etch cycle, the same increase from 20% concentration to 40% leads to a dramatic increase in etch rate, rising from 0.8 nm/min to 2.1 nm/min. The presence of a Ni etch mask atop the micropillar has a minor effect on etch rate, but a dramatic effect on post-etch surface morphology. Following the wet etch, samples with a Ni mask showed much smoother, more vertical sidewalls than samples without the Ni mask. Using optimized etch conditions of 20% AZ400K solution at 90 °C, micropillar sidewall roughness reduced from 33.09 nm to 2.59 nm after immersion in etchant for 60 minutes. This reduction in surface roughness is important for the fabrication of III-Nitride laser diodes, enabling the fabrication of smooth, highly reflective etched mirror faces, avoiding the need for mechanical cleaving. Additionally, this wet etch can lead to improved electrical device performance by reducing or eliminating current leakage pathways, essential for III-Nitride transistors and power electronics.

## EXPERIMENTAL PROCEDURES
### Resource Availability

*Lead Contact*
Further information and requests for resources and reagents should be directed to and will be fulfilled by the Lead Contact, Matthew Seitz (ms3897@rit.edu).

*Materials Availability*
This study did not generate new unique materials.

*Data and Code Availability*
Any data in this study are available from the corresponding author upon reasonable request.

## ACKNOWLEDGMENTS
This work was partially supported by the National Science Foundation, Award No. ECCS 1751675.

## AUTHOR CONTRIBUTIONS
M.S. and J.Z. designed the experiment; J.B. and M.S. conducted the experiments; W.M., N.M., and K.X. collected AFM measurements, M.S, J.B, and B.M. performed data analysis; M.S. was the main contributing author. All authors revised and commented on the manuscript.

## DECLARATION OF INTERESTS
The authors declare no competing interests.